\documentclass[11pt]{article}
\newcommand{\vs}{\vspace{-.15cm}}

\usepackage{amsfonts,amsbsy}
\newcommand{\be}{\begin{equation}}
\newcommand{\ee}{\end{equation}}
\newcommand{\ba}{\begin{eqnarray}}
\newcommand{\ea}{\end{eqnarray}}
\def\Gammabol{{\stackrel{\circ}{\Gamma}}{}}
\def\Abol{{\stackrel{~\circ}{A}}{}}

\def\Rbol{{\stackrel{\circ}{R}}{}}

\def\Gammabol{{\stackrel{\circ}{\Gamma}}{}}

\def\Dbol{{\stackrel{\circ}{\mathcal D}}{}}
\def\nabol{{\stackrel{\circ}{\nabla}}{}}
\def\Lbol{{\stackrel{\circ}{\mathcal L}}{}}
\def\Gammaw{{\stackrel{\bullet}{\Gamma}}{}}

\def\Aw{{\stackrel{~\bullet}{A}}{}}

\def\jw{{\stackrel{~\bullet}{\jmath}}{}}
\def\tw{{\stackrel{\bullet}{t}}{}}

\def\Lw{{\stackrel{\bullet}{\mathcal L}}{}}
\def\Tw{{\stackrel{\bullet}{T}}{}}
\def\Kw{{\stackrel{\bullet}{K}}{}}

\def\Dw{{\stackrel{\bullet}{\mathcal D}}{}}

\def\Sw{{\stackrel{\bullet}{S}}{}}

\def\onehalf{{\textstyle{\frac{1}{2}}}}

\begin{document}
\noindent
{\Large \bf Bringing Together Gravity and the Quanta}
\vskip 0.7cm
\noindent
{\bf R. Aldrovandi, L. C. T. Guillen, J. G. Pereira and K. H. Vu} \\
{\it Instituto de F\'{\i}sica Te\'orica} \\
{\it Universidade Estadual Paulista} \\ 
{\it Rua Pamplona 145} \\ 
{\it 01405-900 S\~ao Paulo, Brazil}
\vskip 0.8cm

\begin{abstract}
\noindent
Due to its underlying gauge structure, teleparallel gravity achieves a separation between  inertial and gravitational effects. It can, in consequence, describe the isolated gravitational interaction without resorting to the equivalence principle, and is able to provide a tensorial definition for the energy-momentum density of the gravitational field. Considering the conceptual conflict between the local equivalence principle and the nonlocal uncertainty principle, the replacement of general relativity by its teleparallel equivalent can be considered an important step towards a prospective reconciliation between gravitation and quantum mechanics.
\end{abstract}

\section{Introduction}

At least at the classical level, the gravitational field exhibits a  quite peculiar property: particles with different masses and different compositions feel it in such a way that all of them acquire the same acceleration and, given the same initial conditions, follow the same path. Such universality of response --- usually referred to as {\it universality of free  fall} --- is the most fundamental characteristic of the gravitational interaction. It is unique, peculiar to gravitation: no other basic interaction of Nature has it. On the other hand, effects equally felt by all bodies were known since long. They are the so called {\it inertial} or {\it fictitious} effects, which show up in non-inertial frames. Examples on Earth are the centrifugal and the Coriolis forces.

Universality of both gravitational and inertial effects was one of the clues used by
Einstein in building up  general relativity, his theory for gravitation. A crucial point of Einstein's description is that it makes no use of the concept of {\em force} for the gravitational
interaction. In fact, instead of acting through a force, gravitation is supposed to produce a {\em curvature} in  spacetime. The gravitational interaction in this case is achieved simply by letting a (spinless) particle to follow a geodesics of the curved spacetime. Notice that no
other kind of spacetime deformation is supposed to exist. Torsion, for example, which
would be another natural spacetime deformation, is assumed to vanish from the very
beginning. In general relativity, therefore, geometry replaces the concept of gravitational force, and the trajectories are determined, not by force equations, but by geodesic equations. The underlying spacetime of this theory is a pseudo-Riemannian space. It is important to emphasize that only an interaction presenting the property of universality can be described  by such a geometrization of spacetime. In the eventual lack of universality, the geometrical description of general relativity would break down. It is also important to observe that universality of free fall is usually identified with the weak equivalence principle. In fact, if all particles move along geodesics, the motion will be independent of their masses --- that is, universal. Nevertheless, in order to allow that  independence, the masses must be canceled out from the equation of motion. Since this cancellation can only be made when the inertial and gravitational masses coincide, this coincidence is  seen as a  statement of the weak equivalence principle.

General relativity and quantum mechanics are not consistent with each other. This conflict stems from the very principles on which these theories take their roots. General relativity, on one hand, is based on the equivalence principle, whose strong version establishes the {\it local} equivalence between gravitation and inertia. It presupposes an {\it ideal} observer on spacetime \cite{ABP02}, represented by a timelike curve which intersects the space-section {\em at a point}. In each space-section, it applies at that intersecting point. The fundamental asset of quantum mechanics, on the other hand, is the uncertainty principle, which is essentially {\it nonlocal}: a test particle is never at a point in space, nor does it  follow a trajectory --- it follows  infinitely many trajectories, each one with a different probability \cite{chiao}. The conflict comes from that observer idealization. In the equation for a curve, gravitation only appears through the Levi--Civita connection, which can be made to vanish all along. An ideal observer can choose frames whose acceleration exactly compensate the effect of gravitation --- which is the strong equivalence principle. A real observer, on the other hand, will be necessarily an object extended in space, consequently intersecting a congruence of curves. Such congruences are described by the deviation equation and will, consequently, detect the true covariant object characterizing the gravitational field, the curvature tensor which cannot be made to vanish. Quantum Mechanics requires real observers, pencils of ideal observers. The inconsistency with the strong equivalence principle is, therefore,  of mathematical necessity. It is not possible, as a consequence, to define a quantum version of this principle \cite{qsep}. On the other hand, the inconsistency of quantum mechanics with the weak equivalence principle is a matter of experimental verification. Although it has passed all experimental tests at the classical level \cite{exp}, there are compelling evidences that the weak equivalence principle might not be true at the quantum level \cite{global}.

Now, in addition to the geometric description of general relativity, gravitation can also be described in terms of a gauge theory \cite{hehl}. In fact, the teleparallel equivalent of general relativity, or teleparallel gravity for short, can be interpreted as a gauge theory for the translation group. In this theory, instead of curvature, torsion is assumed to represent the gravitational field. The corresponding underlying spacetime is, in this case, a Weitzenb\"ock spacetime \cite{weitz}. In spite of this fundamental difference, the two theories are found to yield equivalent classical descriptions of the gravitational interaction \cite{equiva}. Conceptual differences, however, show up. According to general relativity, curvature is used to {\it geometrize} spacetime.Teleparallelism, on the other hand, attributes gravitation to torsion, but in this case torsion accounts for gravitation not by geometrizing the interaction, but by acting as a true {\it force}. As a consequence, there are no geodesics in teleparallel gravity, but only force equations quite analogous to the Lorentz force equation of electrodynamics \cite{paper1}.

One may wonder why gravitation has two different descriptions. This duplicity is related precisely to that peculiarity, universality. Gravitation is, like the other fundamental interactions of nature, described by a gauge theory --- just teleparallel gravity. Universality of free fall, however, makes it possible a second, geometrized description, given by general relativity. As the sole universal interaction, it is the only one to allow also a geometrical interpretation, and two alternative descriptions. One may also wonder why a gauge theory for the translation group, and not for other spacetime group. The reason for this is related to the source of gravitation, that is, energy and momentum. As is well known from Noether's theorem \cite{kopo}, these quantities are conserved provided the physical system is invariant under spacetime translations. It is then natural to expect that the gravitational field be associated to the translation group. This is quite similar to the electromagnetic field, whose source --- the electric four-current --- is conserved due to invariance of the theory under transformations of the unitary group U(1), the gauge group of Maxwell's theory.

Maxwell's gauge theory consistently describes the electromagnetic interaction,  whose non-universal character can be seen from the fact that the Lorentz force depends on the ratio $e/m$ between the electric charge and mass of the particle. This lack of universality precludes a geometric description. On the other hand, although it describes a universal interaction, teleparallel gravity is a gauge theory for the Abelian translation group. Like in Maxwell's theory,  the trajectories are not geodesics, but force equations. As a consequence of this property, teleparallel gravity achieves a separation between inertial from gravitational effects, and is found not to require the equivalence principle to describe the gravitational interaction \cite{wep}. To explore deeper these points will be the basic purpose of these notes. We begin with a review, in the next section, of the fundamentals of teleparallel gravity.

\section{Fundamentals of Teleparallel Gravity}

The mathematical structure of distant parallelism, also referred to as absolute or
teleparallelism, was used by Einstein in the late nineteen twenties, in his  attempt to
unify gravitation with electromagnetism. The crucial idea was the introduction of a 
tetrad field, a field of orthonormal bases of the tangent spaces at each point of the
four-dimensional spacetime. The specification of a tetrad involves sixteen components,
whereas the gravitational field, represented by the spacetime metric, requires only ten 
components. The six additional degrees of freedom ensured  by the tetrad was then supposed by Einstein to represent the electromagnetic field. Due to the local Lorentz invariance of the theory, which eliminates six degrees of freedom, this attempt of unification did not succeed,
but some of the concepts introduced by him remain important up to the present day \cite{uni}. Of course, its present interest concerns the gravitational interaction only, and not any unification purposes.

According to the gauge structure of teleparallel gravity, to each point of spacetime there
is attached a Minkowski tangent space, the ``fiber''  on which the translation (gauge) group acts. We use the Greek alphabet $\mu, \nu, \rho, \dots = 0, 1, 2, 3$ to denote spacetime indices and the Latin alphabet $a, b, c, \dots = 0, 1, 2, 3$ to denote algebraic indices related  to the tangent Minkowski spaces, whose metric is chosen to be $\eta_{a b} = {\rm diag}  (+1,
-1, -1, -1)$. Whenever convenient, the magnitudes related to teleparallel gravity will be denoted with an upper ``$\bullet$'', whereas those corresponding to general relativity will be denoted with an upper ``$\circ$''. As a gauge theory for translations, the fundamental field of teleparallel gravity  is the translational gauge potential $B^a{}_\mu$, a 1-form assuming values in the Lie  algebra of the translation group
\be
B_\mu = B^a{}_\mu \, P_a,
\ee
with $P_a = \partial/ \partial x^a$ the translation generators. The corresponding field strength is
\be
F^a{}_{\mu \nu} = \partial_\mu B^a{}_\nu - \partial_\nu B^a{}_\mu.
\label{tfs1}
\ee
Introducing the trivial --- that is, holonomic --- tetrad
\be
e^a{}_\mu = \partial_\mu x^a,
\ee
which satisfies
\[
\partial_\mu e^a{}_\nu - \partial_\nu e^a{}_\mu = 0,
\]
the field strength can be rewritten in the form
\be
F^a{}_{\mu \nu} = \partial_\mu h^a{}_\nu - \partial_\nu h^a{}_\mu,
\label{tfs2}
\ee
where
\be
h^a{}_\mu = \partial_\mu x^a + B^a{}_\mu
\label{tetrada}
\ee
is an anholonomic tetrad. Since we have chosen its non-gravitational part $\partial_\mu x^a$ to be holonomic, its anholonomy \cite{livro}
\begin{equation}
f^c{}_{a b}  = h_a{}^{\mu} h_b{}^{\nu} (\partial_\nu
h^c{}_{\mu} - \partial_\mu h^c{}_{\nu} )
\label{fcab}
\end{equation}
will come from the gravitational field only, here represented by the translational gauge potential $B^a{}_\mu$. In fact, it is easy to see that
\be
F^a{}_{bc} = - f^a{}_{bc}.
\ee
Although the spacetime metric
\be
g_{\mu \nu} = \eta_{a b} \; h^a{}_\mu \; h^b{}_\nu
\label{gmn}
\ee
does not represent the dynamical variable of the theory, the presence of a translational gauge potential changes its form, which means that teleparallel gravity is not a background-independent field theory. It is enough to observe that, whereas the tangent space indices are raised and lowered with the Minkowski metric $\eta_{ab}$,  spacetime indices are raised and lowered with the spacetime metric $g_{\mu \nu}$.

Teleparallel gravity is characterized by a connection presenting torsion, but no curvature. Its spin connection can always be taken to vanish everywhere in a given frame:
\be
\Aw^{a}{}_{b \mu} \equiv h^{a}{}_{\nu} \left(\partial_{\mu} h_{b}{}^{\nu} +
\Gammaw^\nu{}_{\rho \mu} \, h_b{}^\rho \right) = 0.
\label{wsc}
\ee
This is the so called absolute parallelism condition, from where the theory borrows its name. The corresponding spacetime connection, usually called the Weitzenb\"ock connection, is found to be
\be
\Gammaw^{\rho}{}_{\nu \mu} =
h_{a}{}^{\rho} \partial_{\mu} h^{a}{}_{\nu}.
\label{weit}
\ee
As a simple calculation shows, the field strength is found to coincide with the torsion of the Weitzenb\"ock connection:
\be
F^a{}_{\mu \nu} \equiv \Tw^a{}_{\mu \nu} =
h^a{}_\rho \, (\Gammaw^{\rho}{}_{\nu \mu} -
\Gammaw^{\rho}{}_{\mu \nu}).
\ee

The Lagrangian of teleparallel gravity is \cite{maluf}
\be
\Lw = \frac{h}{4 k^2} \; \Sw_{a}{}^{\mu \nu} \; \Tw^a{}_{\mu \nu},
\label{tela}
\ee
where
\be
\Sw_a{}^{\mu \nu} = h_{a \rho} \, [\Kw^{\mu \nu \rho} - g^{\rho \nu}
\;  \Tw^{\theta \mu}{}_{\theta} + g^{\rho \mu} \; \Tw^{\theta \nu}{}_{\theta}]
\ee
is the superpotential, with
\be
\Kw^{\mu \nu \rho} = {\textstyle \frac{1}{2}} \; (\Tw^{\nu \mu \rho} +
\Tw^{\rho \mu \nu} - \Tw^{\mu \nu \rho})
\ee
the contortion tensor. Considering then the Lagrangian
\begin{equation}
{\mathcal L} = \Lw + {\mathcal L}_m,
\end{equation}
with ${\mathcal L}_m$ representing the Lagrangian of a general matter field, a variation with respect to the gauge field $B^a{}_\rho$ will lead to the teleparallel field equations
\begin{equation}
\partial_\sigma(h \Sw_a{}^{\rho \sigma}) -
k^2 \, (h \jw_{a}{}^{\rho}) = k^2 \, (h {\mathcal T}_{a}{}^{\rho}),
\label{tfe1}
\end{equation}
where
\begin{equation}
\jw_a{}^\rho \equiv - \frac{1}{h} \frac{\partial {\mathcal \Lw}}{\partial h^a{}_\rho} =
\frac{h_a{}^\lambda}{k^2}
\left( \Tw^c{}_{\mu \lambda} \, \Sw_c{}^{\mu \rho} - \frac{1}{4} \, 
\delta_\lambda{}^\rho \,
\Tw^c{}_{\mu \nu} \, \Sw_c{}^{\mu \nu} \right)
\label{emt1bis}
\end{equation}
represents the tensorial form of the gravitational energy-momentum density \cite{prl}, and
\begin{equation}
{\mathcal T}_{a}{}^{\rho} \equiv - \frac{1}{h} \frac{\delta {\mathcal L}_m}{\delta
B^a{}_{\rho}} \equiv - \frac{1}{h} \frac{\delta {\mathcal L}_m}{\delta h^a{}_{\rho}} = - \frac{1}{h}
\left( \frac{\partial {\mathcal L}_m}{\partial h^a{}_{\rho}} -
\partial_\lambda \frac{\partial {\mathcal L}_m}{\partial 
\partial_\lambda h^a{}_{\rho}}
\right)
\label{memt1}
\end{equation}
is the matter energy-momentum tensor. Due to the anti-symmetry of $\Sw_a{}^{\rho \sigma}$
in the last two indices, the total current is conserved as  a consequence of the field
equation:
\begin{equation}
\partial_\rho \left[ h \left( \jw_a{}^\rho + {\mathcal T}_{a}{}^{\rho} \right) 
\right] = 0.
\label{conser0}
\end{equation}

Now, with spacetime indices only, the above field equations acquire the form
\begin{equation}
\partial_\sigma(h \Sw_\lambda{}^{\rho \sigma}) -
k^2 \, (h\, \tw_\lambda{}^\rho) = k^2 \, (h {\mathcal T}_{\lambda}{}^{\rho}),
\label{eqs1}
\end{equation}
where ${\mathcal T}_{\lambda}{}^{\rho} = {\mathcal T}_{a}{}^{\rho} \; h^a{}_\lambda$, and
\begin{equation}
h \, \tw_\lambda{}^\rho = \frac{h}{k^2} \left( \Gammaw^\mu{}_{\nu\lambda} \,
\Sw_{\mu}{}^{\rho \nu} - \frac{1}{4} \, \delta_\lambda{}^\rho \,
\Tw^\theta{}_{\mu\nu} \, \Sw_\theta{}^{\mu\nu} \right)
\label{emt1}
\end{equation}
is the energy-momentum {\it pseudotensor} of the gravitational field. We see clearly from these equations the origin of the connection-term that transforms the tensorial current $\jw_a{}^\rho$ into the energy-momentum pseudotensor $\tw_\lambda{}^\rho$.

The Weitzenb\"ock connection $\Gammaw^\rho{}_{\mu \nu}$ is related to the Christoffel connection $\Gammabol^\rho{}_{\mu \nu}$ of metric $g_{\mu \nu}$  by
\be
\Gammaw^\rho{}_{\mu \nu} = \Gammabol^\rho{}_{\mu \nu} +
\Kw^\rho{}_{\mu \nu} \, .
\label{rela}
\ee
Using this relation, it is possible to show that
\begin{equation}
\Lw = \Lbol - \partial_\mu \left(2 \, h \, k^{-2} \,
\Tw^{\nu \mu}{}_\nu \right),
\end{equation}
where
\begin{equation}
\Lbol = - \frac{\sqrt{-g}}{2 k^2} \; \Rbol
\end{equation}
represents the Einstein--Hilbert Lagrangian of general relativity, with $\Rbol$ the scalar
curvature of the connection $\Gammabol^{\rho}{}_{\mu\nu}$. Up to a divergence, therefore, the teleparallel Lagrangian is equivalent to the Einstein--Hilbert Lagrangian of general relativity.\footnote{It should be remarked that the first-order M{\o}ller's Lagrangian of
general relativity \cite{moller}
\be
\Lbol = \frac{h}{2 k^2} \left(\nabol_{\mu} h^{a \nu} \; \nabol_{\nu}
h_{a}{}^{\mu} - \nabol_{\mu} h_{a}{}^{\mu} \; \nabol_{\nu} h^{a \nu} \right),
\label{mlagra}
\ee
which differs from the Einstein-Hilbert Lagrangian by a total divergence, when rewritten in
terms of the Weitzeb\"ock connection coincides exactly --- that is, without any boundary
term --- with the teleparallel Lagrangian (\ref{tela}). Teleparallel gravity, therefore, can be
considered as fully equivalent with the M{\o}ller's first-order formulation of general
relativity \cite{mollersbook}.} It is important to observe also that, by using the relation (\ref{rela}), the left-hand side of the field equation (\ref{eqs1}) can be shown to satisfy
\begin{equation}
\partial_\sigma(h \Sw_\lambda{}^{\rho \sigma}) -
k^2 \, (h \tw_{\lambda}{}^{\rho}) \equiv
{h} \left({\stackrel{\circ}{R}}_\lambda{}^{\rho} -
\onehalf \, \delta_\lambda{}^{\rho} \;
{\stackrel{\circ}{R}} \right).
\label{ident}
\end{equation}
This means that, as expected due to the equivalence between the cor\-re\-sponding
Lagrangians, the teleparallel field equation (\ref{tfe1}) is equivalent to  Einstein's
field equation
\begin{equation}
{\stackrel{\circ}{R}}_\lambda{}^{\rho} -
\onehalf \, \delta_\lambda{}^{\rho} \,
{\stackrel{\circ}{R}} = k^2 \, {\mathcal T}_{\lambda}{}^{\rho}.
\end{equation}
We see in this way that, in spite of the conceptual differences between teleparallel gravity and general relativity, these theories are found to yield equivalent descriptions of gravitation. From this point of view, general relativity can be considered a complete theory in the sense that it does not need to be generalized through the introduction of torsion \cite{comp}. In fact, observe that the matter energy-momentum tensor appears as the source of curvature in general relativity, and as the source of torsion in teleparallel gravity.

\section{Passage to a General Frame}

In special relativity there is a preferred class of frames: the class of inertial frames.
Similarly, in the presence of gravitation, there will also be a preferred class of frames. These frames can be identified by the fact that, in the absence of gravitation they reduce to the inertial --- that is, holonomic --- frames of special relativity. In these frames, therefore, the coefficient of anholonomy $f^c{}_{ab}$ is related to the gravitational field only, not to inertial effects coming from the frame \cite{reg}. In other words, the first term of the tetrad (\ref{tetrada}), which represents the non-gravitational part of the frame, is holonomic. This is the class of frames commonly used in teleparallel gravity. In these frames, the spin connection vanishes, and the frame anholonomy coincides, up to a sign, with torsion, the tensor that represents the gravitational field in teleparallel gravity:
\be
f^c{}_{ab} = - \Tw^c{}_{ab}.
\label{fimt}
\ee

The above class of frames can thus be said to be purely gravitational in the sense that, in the absence of gravitation, they reduce to the class of inertial frames. Although it assumes a simpler form in these frames, teleparallel gravity can be formulated in any class of frames. Let us then consider a local Lorentz transformation $\Lambda^{a}{}_{b} \equiv \Lambda^{a}{}_{b}(x)$, under which the tetrad changes according to
\be
h'^{a}{}_{\mu} = \Lambda^{a}{}_{b} \; h^{b}{}_{\mu}.
\label{tetratrans}
\ee
A general spin connection, on the other hand, behaves as
\be
A'^{a}{}_{b \mu} = \Lambda^{a}{}_{c} \, A^{c}{}_{d \mu} \, \Lambda_{b}{}^{d} +
\Lambda^{a}{}_{c} \, \partial_{\mu} \Lambda_{b}{}^{c}.
\label{ltsc}
\ee
Since the Weitzenb\"ock spin connection vanishes in a given frame, in the Lorentz-rotated frame it will have the form
\be
\Aw'^{a}{}_{b \mu} = \Lambda^{a}{}_{e} \, \partial_{\mu} \Lambda_{b}{}^{e}. 
\label{ltwsc}
\ee
If the spin connection is zero in a frame, therefore, it will not be zero in any other frame related with the first by a local Lorentz transformation.

Now, in the Lorentz-rotated frame, the tetrad (\ref{tetrada}) will be
\be
h'^a{}_\mu = e'^a{}_\mu + B'^a{}_\mu,
\label{nontetrabis}
\ee
where
\be
e'^a{}_\mu \equiv \Dw'_\mu x'^a = \partial_\mu x'^a + \Aw'^a{}_{b \mu} \, x'^b,
\ee
with $\Aw'^a{}_{b \mu}$ given by Eq.\ (\ref{ltwsc}). Since the translational gauge potential is  Lorentz--covariant,
\be
B'^a{}_\mu = \Lambda^a{}_b \, B^b{}_\mu,
\ee
the inertial effects coming from the frame will be totally included in the first (non-gravitational) term of the tetrad. In fact, whereas the anholonomy of the tetrad (\ref{tetrada}) comes solely from the presence of the gravitational field $B^a{}_\mu$, the anholonomy of the tetrad (\ref{nontetrabis}) includes also inertial effects related to the frame \cite{grg}:
\be
f'^c{}_{ab} = -\,  \Tw'^c{}_{ab} - (\Aw'^c{}_{ab} - \Aw'^c{}_{ba}).
\label{ftaa}
\ee
We see in this way that, even in the absence of gravitation --- an absence here represented by the vanishing of the gravitational field strength $\Tw'^c{}_{ab}$ --- the frame still has an anholonomy, which comes entirely from the inertial properties of the Lorentz-rotated frame. Any theory that has the complete tetrad --- or the metric --- as the field variable, therefore, will necessarily include this additional frame-anholonomy in its fundamental field. This is not the case of teleparallel gravity, whose fundamental field is not the tetrad or the metric, but the isolated translational gauge potential $B^a{}_{\mu}$. In fact, since the frame--related inertial effects  are included in the first term of the tetrad, field $B^a{}_{\mu}$ turns out to represent solely the pure gravitational field. We can then say that teleparallel gravity achieves a separation between inertial and gravitational effects. This is the reason why it does not require the equivalence principle. In what follows, we are going to explore this property in more depth.

\section{Force Equation Versus Geodesics}

Let us consider, in the context of teleparallel gravity, the motion of a spinless particle
of mass $m$ in a gravitational field $B^{a}{}_{\mu}$. The action integral is quite analogous to that of the electromagnetic case,
\begin{equation}
{\mathcal S} =  - m \, c \int_{a}^{b} \left[ u_a \, dx^a +
u_{a} \, B^{a}{}_{\mu} \, dx^{\mu} \right],
\end{equation}
where $u^a = h^a{}_\mu \, u^\mu$ is the anholonomic particle four-velocity, with
\begin{equation}
u^\mu = \frac{d x^\mu}{ds}
\label{ust}
\end{equation}
the holonomic four-velocity, which is written in terms of the spacetime invariant interval
\[
ds = (g_{\mu \nu} dx^\mu dx^\nu)^{1/2}.
\]
In a Lorentz-rotated frame, it assumes the form
\begin{equation}
{\mathcal S} =  - m \, c \int_{a}^{b} \left[ u'_a \, e'^a{}_\mu +
u'_{a} \, B'^{a}{}_{\mu}  \right] dx^{\mu},
\label{acaop1}
\end{equation}
Variation of this action yields
\be
\frac{d u'_a}{ds} - \Aw'^c{}_{ab} \, u'_c \, u'^b = \Tw'^c{}_{ab} \, u'_c \, u'^b.
\label{eqmot3}
\ee
This is the force equation governing the motion of the particle in teleparallel gravity. The right-hand side represents the purely gravitational force, and transforms covariantly under local Lorentz transformations. The fictitious forces coming from the frame non-inertiality, on the other hand, are not covariant, and are taken into account by the connection term appearing in the left-hand side. That the inertial effects are not covariant is clear if we remember that they vanish in an inertial frame. Furthermore, we see from Eq.~(\ref{ltwsc}) that the connection $\Aw'^c{}_{ab}$ depends only on the Lorentz transformation. In teleparallel gravity, therefore, whereas the gravitational effects are described by a covariant force, the inertial effects of the frame remain {\it geometrized} in the sense of general relativity, and are described by a connection.

In general relativity, on the other hand, the inertial and gravitational effects are both geometrized, and described by the same connection. In fact, in that theory the equation of motion is given by the geodesic equation
\begin{equation}
\frac{d u'_a}{d s} - \Abol'^c{}_{ab} \; u'_c \, u'^b = 0,
\label{geode}
\end{equation}
with the spin connection given by
\begin{equation}
\Abol'^c{}_{ab} = \textstyle{\frac{1}{2}} \left( f'^c{}_{ab} + f'_{ab}{}^c - f'_{ba}{}^c \right).
\label{equiva}
\end{equation}
Since $f'^c{}_{ba}$ includes both the gravitational and the inertial effects (see Eq.\ (\ref{ftaa})), these two effects turn out to be mixed in the geodesic equation (\ref{geode}). Using now the relation
\be
\Abol'^c{}_{ab} = \Aw'^c{}_{ab} - \Kw'^c{}_{ab},
\label{relabis}
\ee
as well as the identity
\begin{equation}
-\, \Kw'^c{}_{ab} \; u'_c \, u'^b = \Tw'^c{}_{ab} \; u'_c \, u'^b, 
\label{tuukuu}
\end{equation}
which follows directly from the contortion definition, we see clearly that the geodesic equation (\ref{geode}) is equivalent to the teleparallel force equation (\ref{eqmot3}).  Although equivalent, however, there is a deep difference between these two equations. In the teleparallel approach, the true gravitational effect is extracted from the general relativity spin connection $\Abol'^c{}_{ab}$ and transferred to the right-hand side of the equation of motion, which then becomes a force equation. The inertial effects of the frame remain geometrized through the connection $\Aw'^c{}_{ab}$ appearing in the left-hand side. This means essentially that teleparallel gravity achieves a separation between gravitation and inertia.

\section{Final Remarks}

One of the fundamental problems of gravitation is the conceptual conflict of Einstein's general relativity with quantum mechanics. There are fundamental reasons underneath that inconsistency, essentially related to the very principles on which the theories are based. General relativity, as is well known, is based on the equivalence principle, whose strong version establishes the {\it local} equivalence between gravitation and inertia. Quantum mechanics, on the other hand, is fundamentally based on the uncertainty principle, which is a {\it nonlocal} principle. On this fundamental difference lies one of the roots of the difficulty in reconciling both  theories \cite{chiao}.

Now, it so happens that, in addition to the geometric description of general relativity, gravitation can also be described in terms of a gauge theory for the translation group, called teleparallel gravity. In this theory, instead of curvature, torsion is assumed to represent the gravitational field. In spite of this fundamental difference, the two theories are found to yield equivalent classical descriptions of the gravitational interaction. We may then say that the gravitational interaction can be described {\em alternatively} in terms of curvature, as is usually done in general relativity, or in terms of torsion, in which case we have the so called teleparallel gravity. A crucial property of this theory is that the gravitational effects are taken into account through a true force, whereas the inertial or fictitious effects are geometrized in the sense of general relativity. This can be seen by inspecting the corresponding gravitational coupling rules. Considering a general Lorentz-rotated frame, the coupling prescription of general relativity, as is well known, amounts to replace
\be
\partial_\mu \to \Dbol'_\mu = \partial_\mu - \onehalf \Abol'^{ab}{}_\mu \, S_{ab},
\ee
with $S_{ab}$ the Lorentz generators written in an appropriate representation. In teleparallel gravity, on account of the identity (\ref{relabis}), the coupling prescription is given by \cite{mospe}
\be
\partial_\mu \to \Dw'_\mu = \partial_\mu - \onehalf \Aw'^{ab}{}_\mu \, S_{ab} +
\onehalf \Kw'^{ab}{}_\mu \, S_{ab}.
\ee
We see from these equations that the teleparallel coupling prescription separates the true gravitational effect --- described by contortion --- from the inertial effect --- described by a connection. A crucial point is to observe that curvature, which is the tensor that truly represents the gravitational field in general relativity --- in the sense that it vanishes in the absence of gravitation --- does not appear in the gravitational coupling prescription, nor in the equation of motion of spinless particles. On the other hand, torsion (or equivalently, contortion), which is the tensor that truly represents the gravitational field in teleparallel gravity, does appear in the coupling prescription, and consequently also in the equation of motion of spinless particles. This is a fundamental difference between general relativity and teleparallel gravity, being the responsible for the possibility of separating, in the latter, true gravitational effects from inertia. Furthermore, since torsion can never be made to vanish through an appropriate choice of the frame, the equivalence principle does not apply to this case. Only the inertial effects, which are described by a connection, can be made to vanish in a specific frame.

Another aspect of the same property is the pure-gauge character of the spin connection. It is this property that allows the teleparallel field equation to be written in the potential form (\ref{tfe1}), in which both terms of the left-hand side transform covariantly. Since one of these terms represents the gravitational energy-momentum density, it is consequently a true tensor. It is important to remark that, although Einstein's equation can also be written in the potential form, because its spin connection can never be made to vanish everywhere through an appropriate choice of the frame, its field equation cannot be rewritten in a form in which both terms of the left-hand side are covariant. This means essentially that, in general relativity it is not possible to define a covariant energy-momentum density for the gravitational field \cite{mtw}. This is somehow expected because in general relativity inertial and gravitational effects are always mixed, and since inertial effects are non-covariant, the resulting gravitational energy-momentum density will consequently be non-covariant. As far as in teleparallel gravity it is possible to separate gravitation from inertia, it is also possible to separate the energy-momentum density of the gravitational field --- which, like the teleparallel gravitational force, turns out to be a true tensor --- from that of inertia, which is not tensorial \cite{reg}. An interesting property of the tensorial current (\ref{emt1bis}) is that its trace vanishes identically,
\be
\jw_\rho{}^\rho \equiv h^a{}_\rho \, \jw_a{}^\rho = 0,
\ee
as expected for a massless field. On the other hand, the trace of the pseudotensor
(\ref{emt1}) is found to be proportional to the Lagrangian:
\be
h \, \tw_\mu{}^\mu = - \frac{h}{2 k^2} \; \Sw_{a}{}^{\mu \nu} \; \Tw^a{}_{\mu \nu} = -
2 \, \Lw.
\ee
Similar results hold respectively for the symmetric and the canonical energy-momentum densities of the electromagnetic field \cite{ll}.

Although equivalent to general relativity, the gauge approach of teleparallel gravity is able to separate true gravitational effects --- which is described by a tensor --- from inertial effects --- which is described by a connection. This means that the gravitational interaction in teleparallel gravity is not geometrized, but described by a true force. As a consequence, it turns out not to require the equivalence principle. Replacing general relativity by its teleparallel equivalent, therefore, may be an important step towards a prospective reconciliation of gravitation with quantum mechanics \cite{vaxjo}.

\section*{Acknowledgments}
The authors would like to thank Yu. Obukhov and G. Rubilar for useful discussions. They would like to thank also FAPESP, CNPq and CAPES for financial support.


\end{document}